\documentclass{IEEEtran}
\usepackage[cmex10]{amsmath}
\usepackage{array}
\usepackage{mdwmath}
\usepackage{enumerate}
\usepackage{amssymb}
\usepackage{mdwtab}
\usepackage{graphicx}
\usepackage{stfloats}
\usepackage{cite}
\usepackage{color}
\usepackage{bm}

\newcommand{\tabincell}[2]{\begin{tabular}{@{}#1@{}}#2\end{tabular}}
\usepackage{multirow}
\usepackage{booktabs}
\usepackage[tight,footnotesize]{subfigure}

\begin{document}
\title{\Huge A Distributed Machine Learning-Based Approach for IRS-Enhanced Cell-Free MIMO Networks }
\author{Chen~Chen, \emph{Member, IEEE,}~Sai~Xu, \emph{Member, IEEE,} \\ Jiliang~Zhang, \emph{Senior Member, IEEE,}~and Jie~Zhang, \emph{Senior Member, IEEE}
\thanks{Manuscript received xxx; revised xxx; accepted xxx. Date of publication xxx; date of current version xxx. This work was supported by European Commission’s Horizon 2020 MSCA
Individual Fellowship SICIS under Grant (101032170). The review of this paper was coordinated by xxx. 

Chen Chen (e-mail: C.Chen77@liverpool.ac.uk) is with the Department of Electrical Engineering and Electronics, The University of Liverpool, Liverpool, L69 3GJ, U.K.

Sai~Xu (e-mail: s.xu@sheffield.ac.uk) is with the Department of Electronic and Electrical Engineering, University of Sheffield, Sheffield, S10 2TN, UK. 

Jiliang Zhang (e-mail: zhangjiliang1@mail.neu.edu.cn) is with College
of Information Science and Engineering, Northeastern University, Shenyang,
Liaoning, 314001, China, and was with Department of Electronic and Electrical Engineering, University of Sheffield, Sheffield, S10 2TN, UK and
Ranplan Wireless Network Design Ltd., Cambridge, CB23 3UY, UK during
this work. 

 Jie Zhang (e-mail: jie.zhang@sheffield.ac.uk) is with the
Department of Electronic and Electrical Engineering, University of Sheffield,
Sheffield, S10 2TN, UK, and also with Ranplan Wireless Network Design
Ltd., Cambridge, CB23 3UY, UK.

(Corresponding author: Sai Xu)
}
}
\markboth{IEEE}%
{Shell \MakeLowercase{\textit{et al.}}: Bare Demo of IEEEtran.cls for Journals}
\maketitle

\begin{abstract}
In cell-free multiple input multiple output (MIMO) networks, multiple base stations (BSs) collaborate to achieve high spectral efficiency. Nevertheless, high penetration loss due to large blockages in harsh propagation environments is often an issue that severely degrades communication performance. Considering that intelligent reflecting surface (IRS) is capable of constructing digitally controllable reflection links in a low-cost manner, we investigate an IRS-enhanced downlink cell-free MIMO network in this paper. We aim to maximize the sum rate of all the users by jointly optimizing the transmit beamforming at the BSs and the reflection coefficients at the IRS. 
To address the optimization problem, we propose a fully distributed machine learning algorithm. Different from the conventional iterative optimization algorithms that require a central processing at the central processing unit (CPU) and large amount of channel state information and signaling exchange between the BSs and the CPU, in the proposed algorithm, each BS can locally design its beamforming vectors.  Meanwhile, the IRS reflection coefficients are determined by one of the BSs. Simulation results show that the deployment of IRS can significantly boost the sum user rate and that the proposed algorithm can achieve a high sum user rate with a low computational complexity.

\end{abstract}
\IEEEpeerreviewmaketitle
\section{Introduction}
{With the surge of data traffic in the fifth generation (5G)-and-beyond wireless communication networks, small-cell base stations and large-scale antenna arrays have been increasingly deployed}~\cite{bjornson2016deploying}. {The colocated multiple-input multiple-output (MIMO) architecture adopted by these BSs, however, leads to severe intercell interference for cell-edge users \cite{9791134}.}  In this context, cell-free MIMO has emerged as an attractive solution {to eliminating intercell interference. In such a new network architecture,} all the BSs cooperatively serve the users {occupying} the same time-frequency resources by exploiting the aggressive spatial user multiplexing \cite{7827017, 9064545}. {Connected to the central processing unit (CPU) through front-haul links, the BSs simultaneously send their data streams to the served users.} Thereby, cell-free MIMO networks experience no cell boundaries and achieve uniform coverage at high spectral efficiency. However, cell-free MIMO networks {often} suffer from a low signal-to-noise ratio (SNR) {due to the} harsh propagation environment, {e.g.} urban areas with dense buildings \cite{9665300}. {Although deploying more multi-antenna BSs can alleviate the issue, conventional large-scale antenna arrays result in skyrocketing hardware and power costs as well as signal processing complexity.}

{As one of the most cutting-edge techniques in recent years, intelligent reflecting surface (IRS) has shown a huge potential to address the issue of harsh propagation, in a cost- and energy-efficient manner \cite{wu2021intelligent, 9311936, 9140329}}. To be specific, IRS is a uniform planar array consisting of a large number of programmable passive reflection elements whose amplitudes and phases can be adaptively adjusted. Accordingly, IRS is capable of achieving smart radio environment by creating favorable propagation conditions without using active radio-frequency chains/amplifiers. Recently, IRS has found wide applications in wireless communications \cite{guo8982186, pan9090356, dong9159923, xu9667508}.  In \cite{guo8982186}, the authors investigated an IRS-assisted single-cell downlink multi-user MIMO network, where the weighted sum-rate (WSR) of all users was maximized by optimizing the transmit beamforming and IRS phase shifts. In \cite{pan9090356}, the authors considered a multi-cell MIMO scenario and jointly optimized the transmit beamforming at each BS and the IRS phase shifts {to} maximize the WSR by utilizing a block coordinate descent algorithm. In \cite{dong9159923}, the authors targeted at maximizing the secrecy rate for an IRS-aided MIMO system with a multi-antenna eavesdropper. In \cite{xu9667508} and \cite{10017389}, IRS backscatter-assisted downlink and uplink  multi-cell communication networks were investigated, respectively, and the beamforming vectors at the power beacon and the IRSs were optimized to maximize the WSR.

There have been several works integrating the IRS into cell-free MIMO networks \cite{globecom9685730, yu19286726, Noh9716138, yu29352948, huang9298843}. In \cite{globecom9685730}, tight approximation expressions of the outage probability and achievable rate were analytically derived. The numerical results revealed that the performance of cell-free communication networks can be enhanced by deploying an IRS. In \cite{yu19286726}, multiple IRSs were employed to assist a single-cell communication system, where the transmit beamforming and the IRS-based beamforming were optimized to maximize the sum user rate. In \cite{Noh9716138}, an optimization problem was formulated to maximize the minimum achievable rate. In particular, the transmit beamforming was optimized based on instantaneous channel state information (CSI) and the IRS phase shifts were optimized based on the statistical CSI. In \cite{yu29352948}, the authors focused on the energy efficiency maximization problem. However, the aforementioned works adopted centralized optimization algorithms that require a large amount of CSI and signaling exchanges between the BSs and the CPU, which limits the system scalability. The authors in \cite{huang9298843} proposed a decentralized optimization algorithm for cooperative beamforming in IRS-assisted cell-free MIMO networks. Although no CSI exchange is required, the optimization variables need to be shared among neighboring BSs during the iterations of the algorithm. Up to now, a fully distributed algorithm that can jointly optimize the  transmit beamforming and IRS reflection coefficients without CSI and signaling exchanges is still missing. Moreover, these works rely on iterative optimization algorithms that would incur {an} immense computational complexity.

As a key enabling technology for 5G-and-beyond wireless communication systems, machine learning has potential to solve mathematically intractable non-convex optimization problems \cite{Survey1, Magzine2}. It has been used in IRS-assisted single-cell communication networks \cite{feng8968350, huang9110869, yang9206080, tao9427148}. In \cite{feng8968350}, the authors investigated a single-user multiple-input single-output scenario, where the SNR is maximized by optimizing IRS phase shifts using deep reinforcement learning (DRL). The authors in \cite{huang9110869} further studied a multi-user scenario and applied DRL to jointly optimize the transmit beamforming matrix at the BS and the phase shifts at the IRS.   In \cite{yang9206080}, DRL was used to design secure beamforming  with the worst-case secrecy rate and data rate constraints. In \cite{tao9427148}, the  beamformers at the BS and the IRS in a single-cell dowlink communication network were  designed directly based on received pilots using a graph neural network (GNN).

To the best of our knowledge, machine learning has not been exploited in IRS-assisted cell-free MIMO networks. In this paper, we propose a novel distributed machine learning algorithm to jointly optimize the beamforming vectors at the BSs and the reflection coefficient
vector at the IRS.  Our
major contributions are summarized as follows: 
\begin{itemize}
	\item We investigate an IRS-enhanced cell-free MIMO network and formulate a joint transmit beamforming and reflect beamforming optimization problem aiming to maximize the sum user rate. Then we develop an centralized alternative optimization (AO)-based baseline algorithm to solve the optimization problem.
	\item We novelly propose a fully distributed machine learning algorithm where each BS determines its own beamforming vectors using a GNN based on its local CSI. The IRS reflection coefficients can be determined by only one of the BSs. Our proposed distributed machine learning framework does not require any CSI exchange, and thus is front-haul friendly. 
	\item Different from conventional optimization algorithms that rely on iterative optimization, the proposed algorithm can directly obtain the transmit and reflect beamforming vectors as the outputs of the well trained GNNs. 
 We provide extensive simulation results to demonstrate the effectiveness of the proposed  distributed machine learning approach in terms of boosting the sum user rate.
\end{itemize}

The rest of the paper is organized as follows. Section \ref{sec:system_model} presents the system model and formulates the optimization problem. An AO-based baseline solution is introduced in Section \ref{sec:baseline}. In Section \ref{sec:ML}, a distributed machine leaning-based algorithm for joint design of BS transmit beamforming and IRS reflection coefficients is illustrated.  Simulation results are provided in Section \ref{sec:simulation}, whereas main conclusions are drawn in Section \ref{sec:conclusions}.

Notations: In this paper, scalars, vectors and matrices are represented by italic letters, boldface lower-case letters,
and boldface upper-case letters, respectively.  $(\cdot)^{T}$ and $(\cdot)^{H}$ and $\text{Tr}(\cdot)$ are the transpose, conjugate transpose and trace
operations, respectively. 
$[\mathbf{A}]_{i,j}$ denotes the $(i,j)$-th element of a matrix $\mathbf{A}$. 
$\text{vec}(\mathbf{A})$ is the vectorization of a matrix $\mathbf{A}$ and
$\text{diag}(\mathbf{a})$ is a diagonal matrix whose diagonal elements are the corresponding elements in vector $\mathbf{a}$. $\mathcal{CN}(\mu,\sigma^2)$ stands for a circularly symmetric complex Gaussian  distribution with mean $\mu$ and variance $\sigma^2$. $\Re\{a\}$ and  $\Im\{a\}$ denote the real part and imaginary part of a complex number $a$, respectively. $\mathbb{C}^{M\times N}$ and $\mathbb{R}^{M\times N}$ represent the space of complex-valued and real-valued matrices, respectively. $\otimes$ is the Kronecker product. 

\section{System Model and Problem Formulation}
\label{sec:system_model}
\subsection{System Model}
\begin{figure}[!t]
\centerline{\includegraphics[width= 3.2in]{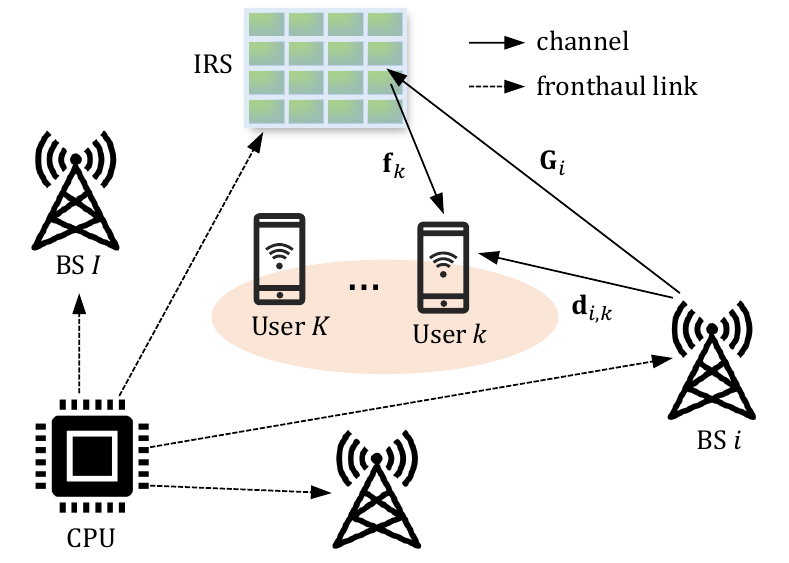}}
\caption{An illustration of the IRS-enhanced downlink cell-free MIMO network.}
\label{fig:system}
\end{figure}
Fig. \ref{fig:system} shows an IRS-enhanced downlink cell-free MIMO network, consisting of $I$ multi-antenna BSs with absolutely identical hardware configuration, $K$ single-antenna users and an IRS. Each of the BSs is equipped with $M$ antennas and the IRS has $L$ passive reflecting elements. The index sets of BSs, users, and IRS elements are denoted by $\mathcal{I}= \{1,2,\cdots,I\}$, $\mathcal{K}= \{1,2,\cdots,K\}$ and $\mathcal{L}= \{1,2,\cdots, L\}$, respectively. It is assumed that only the phase shift is adjustable for each element of the IRS. The reflection coefficient vector of the IRS is given by $\textbf{v} =\left[e^{j\theta_{1}}, e^{j\theta_{2}}, \cdots, e^{j\theta_{L}}\right]$, where $\theta_{l}\in [0,2\pi)$, $l \in \mathcal{L}$. Under the control of the central processing unit (CPU), the BSs  collaborate to send the information to the users while the IRS is reasonably deployed to enhance the communication. \par
Let $\textbf{d}_{i,k} \in \mathbb{C}^{M \times 1}$, $\textbf{G}_{i}\in \mathbb{C}^{L \times M}$ and $\textbf{f}_{k}\in \mathbb{C}^{L\times 1}$ represent the channels from user $k$ to BS $i$, from BS $i$ to the IRS and from user $k$ to the IRS, respectively. The superposed channel from BS $i$ to user $k$ is given by
\begin{align}
\textbf{h}_{i,k}^H = \textbf{d}_{i,k}^H +  \textbf{f}_{k}^H \text{diag}(\textbf{v}^H){\textbf{G}_{i}}.
\end{align}
In the cell-free system, the data requested by each user is simultaneously sent by all the BSs. Thus, the received signal at user $k$ is given by 
\begin{align}
 y_{k} = \underbrace{\sum_{i\in \mathcal{I}} {\textbf{h}}_{i,k}^H \textbf{w}_{i,k}{s}_{k}}_{\text{the desired signal}} + \underbrace{\sum_{i\in \mathcal{I}}\sum_{k'\in \mathcal{K}, k'\ne k} {\textbf{h}}_{i,k}^H \textbf{w}_{i,k'} {s}_{k'}}_{\text{the interference signals}} + n_k,
\end{align}
where $\textbf{w}_{i,k}\in \mathbb{C}^{M\times 1}$ is the beamforming vector from BS $i$ to user $k$, $s_{k}$ is the data symbol for user $k$, and $n_{k}\sim \mathcal{CN}(0, \delta^{2})$ is the additive white Gaussian noise. For brevity, we define $\textbf{h}_k = \left[\textbf{h}_{1,k}; \textbf{h}_{2,k}; \cdots; \textbf{h}_{I,k}  \right]\in \mathbb{C}^{IM \times 1}$, $\textbf{d}_k = \left[\textbf{d}_{1,k}; \textbf{d}_{2,k}; \cdots; \textbf{d}_{I,k}  \right]\in \mathbb{C}^{IM \times 1}$, $\textbf{G} = \left[\textbf{G}_1, \textbf{G}_2, \cdots, \textbf{G}_I \right]\in \mathbb{C}^{L \times IM}$ and $\textbf{w}_k = \left[\textbf{w}_{1,k}; \textbf{w}_{2,k}; \cdots; \textbf{w}_{I,k}  \right]\in \mathbb{C}^{IM \times 1}$. Then, ${\textbf{h}}_{k}^H = \textbf{d}_{k}^H +  \textbf{f}_{k}^H \text{diag}(\textbf{v}^H) \textbf{G}$. The received signal at user $k$ is further given by 
\begin{align}
 y_{k} = \textbf{h}_k^H \textbf{w}_k s_k + \sum_{k'\in \mathcal{K}, k' \ne k} \textbf{h}_k^H \textbf{w}_{k'} s_{k'} + n_k.
\end{align}
Accordingly, the achievable data rate of user $k$ is given by
\begin{align}
 R_{k}(\textbf{w}_k, \textbf{v}) = \text{log}_{2}\left(1 + \frac{\left| {\textbf{h}}_k^H \textbf{w}_k \right|^{2}}{ \sum_{k'\in \mathcal{K}, k' \ne k}  \left| {\textbf{h}}_k^H \textbf{w}_{k'} \right|^{2} + \delta^{2}}\right).
\end{align}
\subsection{Problem Formulation}
This paper aims to optimize the beamforming vectors $\mathcal{W}=\{\textbf{w}_{i,k}|\forall i \in \mathcal{I}, \forall k \in \mathcal{K}\}$ at the BSs and the reflection coefficient vector $\textbf{v}$ at the IRS to maximize the sum rate of all the users. Mathematically, the optimization problem is formulated as 
\begin{align}
(\mathcal{P}0)~\mathop{\max_{\mathcal{W}, \textbf{v}}}~& \sum_{k\in \mathcal{K}}R_{k}(\textbf{w}_k, \textbf{v})  \label{P}   \\
      s.t.~          
      & \sum_{k\in\mathcal{K}}\text{Tr} \left( \textbf{w}_{i,k} \textbf{w}_{i,k}^H \right) \le P_{\text{max}}, i\in \mathcal{I}, \tag{\ref{P}{a}}  \\
      &   \left|\textbf{v}_{l}\right| = 1, l \in \mathcal{L}, \label{c1b} \tag{\ref{P}{b}}
\end{align}
where $P_{\text{max}}$ is the maximum transmit power per BS.
\section{Baseline Solution}
\label{sec:baseline}
In this section, we develop an baseline solution to optimize the transmit beamforming vectors where the IRS  reflection coefficients are randomly configured. 
We first transform the problem $(\mathcal{P}0)$ into a tractable form using fractional  programming, and then employ the method of AO to solve it.\par
\subsection{Problem Transformation}
Using Lagrangian dual transform to tackle the logarithm functions and introducing auxiliary variable $\alpha_{k}$, the objective function is rewritten as
\begin{align}
 &  \sum_{k\in\mathcal{K}}  R_{k}(\textbf{w}_k, \textbf{v}) =\underset{\alpha_{k} \geq 0} \max ~  \sum_{k\in\mathcal{K}}  \log \left( 1 + \alpha_{k} \right) - \alpha_{k}  + \frac{\left( 1 + \alpha_{k} \right) \gamma_{k} }{ 1 + \gamma_{k}}, 
\end{align}
where $\gamma_{k}$ denotes the signal-to-interference-and-noise-ratio (SINR) at user $k$. Introducing an auxiliary variable ${\beta}_{k}$,
\begin{align}
  &\quad  \sum_{k \in \mathcal{K}}  \frac{\left( 1 + \alpha_{k} \right) \gamma_{k} }{ 1 + \gamma_{k}} =  \sum_{k \in \mathcal{K}}  \frac{ \left( 1 + \alpha_{k} \right)  | \mathrm{A}_{k} |^2 }{ \mathrm{B}_{k} }    \nonumber\\
&=  \sum_{k\in \mathcal{K}}   2 \sqrt{ 1 + \alpha_{k} } \Re \{ {\beta}_{k}^\ast \mathrm{A}_{k}\} -  |{\beta}_{k}|^2 \mathrm{B}_{k},
\end{align}
where $\mathrm{A}_{k}$ and $\mathrm{B}_{k}$ are respectively given by
%
%
\begin{align}
\mathrm{A}_{k} &= {\textbf{h}}_k^H {\textbf{w}}_k = \sum_{i\in \mathcal{I}} {\textbf{h}}_{i,k}^H \textbf{w}_{i,k},  \\
\mathrm{B}_{k} &= \delta^{2} + \sum_{k'\in \mathcal{K}}  \left| {\textbf{h}}_k^H \textbf{w}_{k'} \right|^{2} = \delta^{2} + \sum_{k'\in \mathcal{K}}  \left| \sum_{i\in \mathcal{I}} {\textbf{h}}_{i,k}^H \textbf{w}_{i,k'} \right|^{2}.
\end{align}

We consider zero-forcing (ZF) beamforming with power allocation (PA) as the transmit beamformer to eliminate inter-user interference and achieve near-optimal solution \cite{huang2019reconfigurable, yu29352948}.  The global channel information of all the users can be expressed as $\textbf{H} = \left[\textbf{h}_{1}, \textbf{h}_{2}, \cdots, \textbf{h}_{K}  \right]\in \mathbb{C}^{IM \times K}$, and the global ZF precoding matrix is computed by 
\begin{align}
\label{eq:ZF}
\widetilde{\textbf{W}} =  \textbf{H}\left(\textbf{H}^{H}\textbf{H}\right)^{-1},
\end{align}
where $\widetilde{\textbf{W}} \in \mathbb{C}^{IM \times K}$. Then we have $\textbf{w}_{i,k}=\sqrt{P_{i,k}}\widehat{\textbf{w}}_{i,k}$, where $P_{i,k}$ represents the transmit power allocated to the $k$-th user at the $i$-th BS and $\widehat{\textbf{w}}_{i,k}= \widetilde{\textbf{w}}_{i,k}/||\widetilde{\textbf{w}}_{i,k}||$ with $\widetilde{\textbf{w}}_{i,k}=[\widetilde{\textbf{W}}]_{M(i-1)+1:Mi,k}$ is the normalized beamforming vector from BS
$i$ to user $k$. Accordingly, the new objective function is given by
\begin{align}
f(P_{i,k}, &\alpha_k, {\beta}_k) = \underset{\alpha_{k} \geq 0} \max ~  \sum_{k \in \mathcal{K}}  \log \left( 1 + \alpha_{k} \right) - \alpha_{k}  \nonumber\\
& + \sum_{k\in \mathcal{K}}   2 \sqrt{ 1 + \alpha_{k} } \Re \{ {\beta}_{k}^\ast \mathrm{A}_{k}\} -    |{\beta}_{k}|^2 \mathrm{B}_{k}.
\end{align}
With the IRS  reflection coefficient vector $\textbf{v}$ randomly configured, the problem $(\mathcal{P}0)$ is reformulated as
\begin{align}
(\mathcal{P}1) ~\underset{ P_{i,k}, \alpha_{k}, {\beta}_{k}} \max \quad &   f(P_{i,k},  \alpha_k, {\beta}_k), \label{P1}\\
s.t. \quad
 &  \alpha_{k} \geq 0, ~k \in \mathcal{K}, \tag{\ref{P1}{a}}\\
 & \sum_{k\in\mathcal{K}} P_{i,k} \le P_{\text{max}}, i\in \mathcal{I}. \tag{\ref{P1}{b}}
\end{align}
However, this problem is still difficult to solve directly. In view of this, we will apply the method of AO to solve it in the following subsection.
\subsection{Problem Optimization}
This subsection will develop an alternate method to address the optimization problem by two steps. In the first step, the variables $\alpha_{k}$ and ${\beta}_{k}$ are optimized with $P_{i,k}$ fixed. In the second step, the variable $P_{i,k}$ is optimized with $\alpha_{k}$ and ${\beta}_{k}$ fixed.

\vspace{0.1cm}
\noindent \emph{Step-1):} Optimizing $\alpha_{k}$ and ${\beta}_{k}$ \par
\vspace{0.1cm}

With $P_{i,k}$ given, we separately take the derivative of $\alpha_{k}$ and ${\beta}_{k}$ to find the optimal $\alpha_{k}^\circ$ and ${\beta}_{k}^\circ$. It is deduced that the optimal $\alpha_{k}^\circ$ and ${\beta}_{k}^\circ$ are given by
\begin{align}
& \frac{\partial  f(P_{i,k}, \alpha_k, {\beta}_k) }{\partial \alpha_{k}} = 0,    \\
&\frac{\partial   f(P_{i,k}, \alpha_k, {\beta}_k) }{\partial {\beta}_k } = 0. 
\end{align}
It is derived that the optimal $\alpha_{k}^\circ$ and ${\beta}_{k}^\circ$ are respectively given by
\begin{align}
\alpha_{k}^\circ &=  \gamma_{k}, \\
{\beta}_{k}^\circ &=  \frac{\sqrt{ 1 + \alpha_{k}^\circ }  \mathrm{A}_{k} }{\mathrm{B}_{k}}. 
\end{align}
\vspace{0.1cm}
\noindent \emph{Step-2):} Optimizing $P_{i,k}$ \par
\vspace{0.1cm}
With $\alpha_{k}$ and ${\beta}_{k}$ given, the objective function can be simplified as
%
\begin{align}
&\underset{P_{i,k}} \max \quad  f(P_{i,k}, \alpha_k, {\beta}_k)  \nonumber \\
\Longleftrightarrow & \underset{P_{i,k}} \max ~ \sum_{k} 
 2 \sqrt{ (1 + \alpha_{k})} \Re \{ {\beta}_{k}^\ast  \sum_{i\in \mathcal{I}}  \sqrt{P_{i,k}} {\textbf{h}}_{i,k}^H \widehat{\textbf{w}}_{i,k}\}   \nonumber \\
 & \quad\quad\quad\quad\quad -    |{\beta}_{k}|^2 \sum_{k'\in \mathcal{K}}   \left|  \sum_{i\in \mathcal{I}}  \sqrt{P_{i,k'}} {\textbf{h}}_{i,k}^H \widehat{\textbf{w}}_{i,k'} \right|^{2}.
\end{align}
%
%
Thus, the problem is equivalently reformulated as
\begin{align}
(\mathcal{P}2) ~ \underset{ \{P_{i,k}\}} \max \quad &   \sum_{k} 
 2 \sqrt{ (1 + \alpha_{k})} \Re \{ {\beta}_{k}^\ast  \sum_{i\in \mathcal{I}}  \sqrt{P_{i,k}} {\textbf{h}}_{i,k}^H \widehat{\textbf{w}}_{i,k}\}   \nonumber \\
 &    -    |{\beta}_{k}|^2 \sum_{k'\in \mathcal{K}}   \left|  \sum_{i\in \mathcal{I}}  \sqrt{P_{i,k'}} {\textbf{h}}_{i,k}^H \widehat{\textbf{w}}_{i,k'} \right|^{2},  \label{P2}\\
s.t. \quad
 &  \sum_{k\in\mathcal{K}} P_{i,k} \le P_{\text{max}}, i\in \mathcal{I}.\tag{\ref{P2}{a}}
\end{align}
This problem is convex over $P_{i,k}$ and easy to solve. \par

\subsection{Complexity Analysis}
The computational complexity of ZF beamforming in (\ref{eq:ZF}) is $\mathcal{O}(IMK^2)$ \cite{bjornson2017massive, nayebi2017precoding}.
The total computational complexity of the baseline solution is approximately given by
\begin{align}
C_\text{base} = T_\text{ite} \left[ \frac{1}{3} \mathcal{O}\left( (IM)^3 \right) + 2 \mathcal{O}\left( (IM)^2 \right) \right] + \mathcal{O}(IMK^2)
\end{align}
where $T_\text{ite}$ represents the iteration number for the alternative optimization in Section III-B.
\section{Proposed Distributed Machine Learning Algorithm}
\label{sec:ML}
The baseline solution requires iterative optimization, which incurs a high computational complexity. Moreover, it is a centralized algorithm that relies on global CSI information. The optimal beamforming vectors are determined by the CPU and then transmitted to the BSs, resulting in {a} heavy front-haul overhead.

In this section, we aim to develop a distributed machine learning algorithm where each BS uses an independent GNN to design the local beamforming vectors according to its local CSI information. Moreover, the reflection coefficients of the IRS are controlled by only one of the BSs locally. The structure of the proposed distributed machine learning algorithm will be elaborated in the following subsections.

\subsection{Preliminaries of GNN}
Graphs are used to characterize non-Euclidean data. A typical graph is expressed as $\mathcal{G}=\left(\mathcal{V}, \mathcal{E}\right)$, where $\mathcal{V}$ and $\mathcal{E}$ denote the sets of nodes and edges, respectively \cite{TNNLS1}. A node is denoted by $v_{k}\in \mathcal{V}$ and an edge is denoted by $(v_{k},v_{j})\in \mathcal{E}$. The connections between the nodes are represented by an adjacency matrix $\textbf{A}$, where $[\textbf{A}]_{k,j}=1$ if edge $(v_{k},v_{j})\in \mathcal{E}$, otherwise, $[\textbf{A}]_{k,j}=0$. For an undirected graph, the adjacency matrix is symmetric, and the main diagonal elements are all zero. Each node has its own attribute feature. In this paper, the feature of $v_{k}$ is characterized by a node feature vector $\textbf{x}_{k}$. We adopt spatial graph convolutional networks to update the node feature vectors by aggregating information from their neighbour nodes. Under the message passing neural network framework \cite{gilmer2017neural}, the node feature vector of  $v_{k}$ in the $n$-th iteration is given by
\begin{align} 
\textbf{x}_{k}^{n} &= \Omega^{n}\left(\Delta^{n}\left(\left\{\textbf{x}_{k^{'}}^{n-1} |  \forall v_{k^{'}}\in \mathcal{N}(v_{k})\right\}\right), \textbf{x}_{k}^{n-1}\right),
\end{align}
where $\Delta^{n}$ is the set aggregation function and $\Omega^{n}$  is the combination function ({For more details, please kindly refer to Section \ref{sec:structure of gNNs}}). The parameters of $\Delta^{n}$ and $\Omega^{n}$ are shared among all the nodes, and thus GNNs  are scalable to varying node sizes \cite{he9618652}. On the other hand, the conventional neural networks can only be applied to a particular number of nodes. In our considered IRS-enhanced cell-free MIMO network, the number of users may vary over time. In this light, the number of conventional neural networks required is equal to {that} of possible users. This observation motivates us to adopt GNNs.

\subsection{Graph Representation of the IRS-Enhanced Cell-Free MIMO Network}
\label{sec:graph representation}
\begin{figure}[!t]
\centerline{\includegraphics[width= 3.2in]{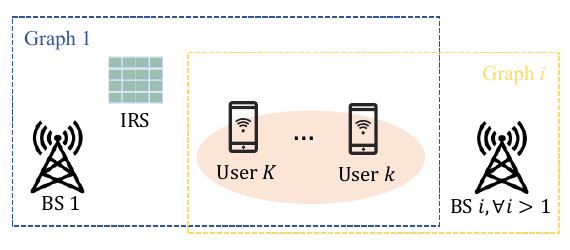}}
\caption{An illustration of graph modelling of the IRS-enhanced cell-free network.}
\label{fig:graph_construction}
\end{figure}

 We will show in Section \ref{sec:numerical_results} that it  does not matter which BS determines the IRS reflection coefficients. Without loss of generality, we assume that BS $1$ determines  the IRS reflection coefficients in this paper. As shown in Fig. \ref{fig:graph_construction}, we construct a graph for each local network. Specially, in addition to its local beamforming vectors, BS $1$ also determines the configuration of the IRS reflection coefficients.  Hence, graph $1$ is constructed by BS $1$, the $K$ users and the IRS. The other BSs except BS $1$ only determine their own local beamforming vectors. Accordingly, graph $i$, $\forall i>1$, is constructed by BS $i$ and the $K$ users. 

\begin{figure}[!t]
\centering
\subfigure[]{
\begin{minipage}[t]{0.5\linewidth}
\centering
\includegraphics[width= 1.7in]{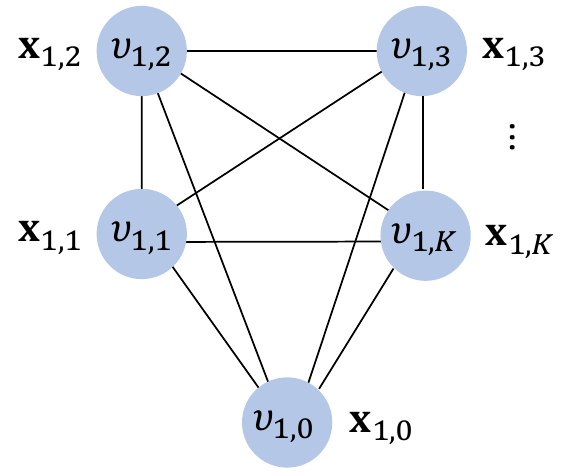}
\end{minipage}%
}%
\subfigure[]{
\begin{minipage}[t]{0.5\linewidth}
\centering
\includegraphics[width= 1.7in]{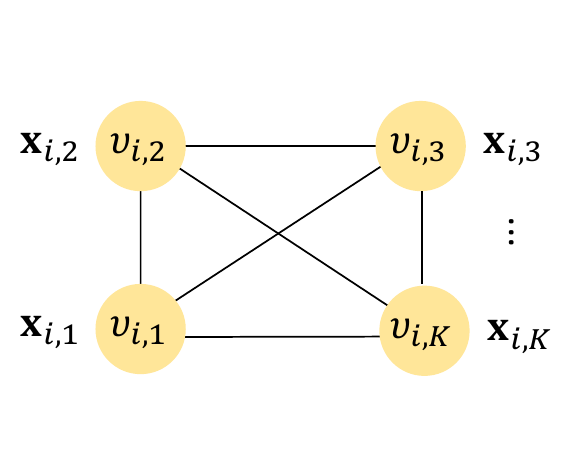}
\end{minipage}%
}%
\centering
\caption{(a) An illustration of graph 1.  (b) An illustration of graph $i$,  $\forall i > 1$.}
\label{fig:graph_representation}
\end{figure}

The structures of the graphs are shown in Fig. \ref{fig:graph_representation}, where $v_{i,k}$ denotes node $k$ of graph $i$ and $\textbf{x}_{i,k}$ denotes the  node feature vector of $v_{i,k}$. Graph $1$ is defined as $\mathcal{G}_{1}=\left(\mathcal{V}_{1}, \mathcal{E}_{1}\right)$, where $\mathcal{V}_{1}=\left\{v_{1,0}, v_{1,1},\cdots, v_{1,K}\right\}$ is the set of nodes and $\mathcal{E}_{1}$ is the set of edges in graph $1$. More specifically, $v_{1,0}$ represents the IRS and $v_{1,k}, \forall k\in \mathcal{K}$, represents user $k$. Similarly, graph $i$, $\forall i>1$, is defined as $\mathcal{G}_{i}=\left(\mathcal{V}_{i}, \mathcal{E}_{i}\right)$, where $\mathcal{V}_{i}=\left\{v_{i,1},\cdots, v_{i,K}\right\}$ is the set of nodes with  $v_{i,k}$ representing user $k$ and $\mathcal{E}_{i}$ is the set of edges in graph $i$. Each graph is constructed as a undirected graph, where each node is connected to all other nodes, i.e., an edge exists between any two nodes. As such, the adjacency matrix of $\mathcal{G}_{i}, i\in \mathcal{I}$, denoted by $\textbf{A}_{i}$, has $[\textbf{A}_{i}]_{k,j}=1, \forall k\ne j$.

\subsection{Structure of GNNs}
\label{sec:structure of gNNs}
The aim of the GNNs is to obtain efficient node feature vectors that contain enough information to design the BS beamforming vectors and IRS reflection coefficient vector.  The overall structure of the GNNs used is shown in Fig.  \ref{fig:GNNs} and is detailed as follows:

\begin{figure*}[!t]
\centering
\subfigure[The structure of GNN 1.]{
\begin{minipage}[t]{1\linewidth}
\centering
\includegraphics[width= 6in]{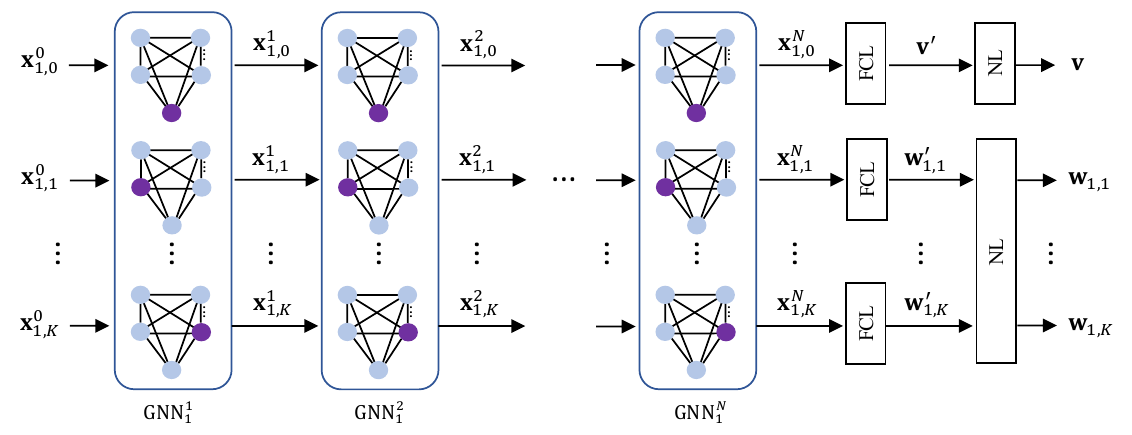}
\end{minipage}
} \\
\subfigure[The structure of GNN $i$, $\forall i>1$.]{
\begin{minipage}[t]{1\linewidth}
\centering
\includegraphics[width= 6in]{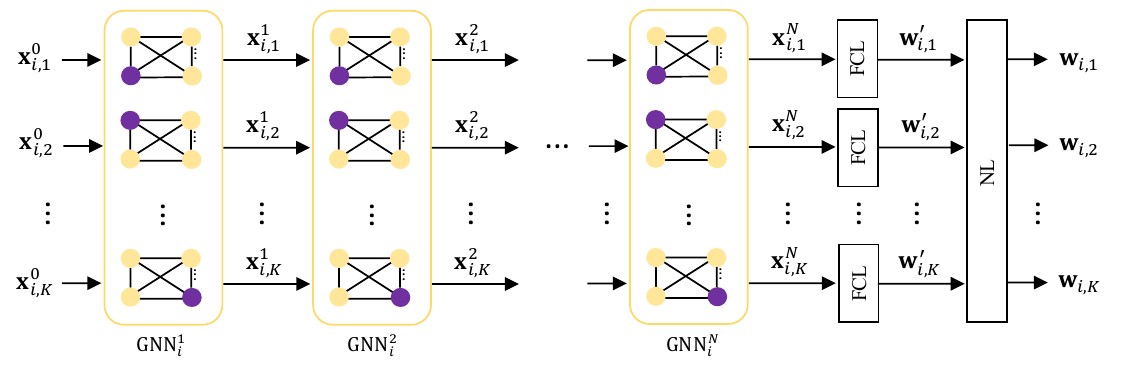}
\end{minipage}
}
\centering
\caption{The proposed structure of the GNNs. Each purple node represents an update to the node feature vector.}
\label{fig:GNNs}
\end{figure*}

\subsubsection{Input} The initial node feature vector taking the node input feature of $v_{i,k}$ is denoted by $\textbf{x}_{i,k}^{0}$. The input feature of user node $v_{i,k}, \forall i\in \mathcal{I}, k\in \mathcal{K}$, is the vectorized CSI from BS $i$ to user $k$ given by
\begin{align}
\textbf{x}_{i,k}^{0}\!=\![\Re\left\{\textbf{d}_{i,k}^{T}\right\}, \Im\left\{\textbf{d}_{i,k}^{T}\right\}, \text{vec}(\Re\left\{\textbf{C}_{i,k}\right\})^{T}, \text{vec}(\Im\left\{\textbf{C}_{i,k}\right\})^{T}]^{T},
\end{align}
where $\textbf{C}_{i,k} = \textbf{G}_{i}\text{diag}\left(\textbf{f}_{k}\right)$ is the cascaded channel from BS $i$ to user $k$ through the reflection of the IRS. The CSI information is separated by real and imaginary components as neural networks only support real-valued input. The dimension of $\textbf{x}_{i,k}^{0}$ is $2M(L+1)$.

Since the IRS contains channel information of all the user channels, we design the input feature of the IRS node as the mean of user input features. Recall that the IRS node is only included in graph $1$, and thus the input feature of the IRS node is given by
\begin{align}
\textbf{x}_{1,0}^{0} = F_{\text{mean}}\left(\textbf{x}_{1,1}^{0}, \textbf{x}_{1,2}^{0}, \cdots, \textbf{x}_{1,K}^{0}\right),
\end{align}
where $F_{\text{mean}}(\cdot)$ is the element-wise mean function. 

\subsubsection{Graph  convolutions}
\begin{figure}[!t]
\centerline{\includegraphics[width= 3.4in]{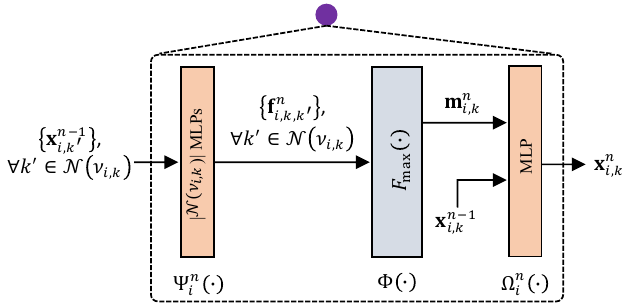}}
\caption{The proposed node update rule.}
\label{fig:network_node}
\end{figure}

The initial node feature vectors are updated by $N$ layers of graph  convolutions. In our proposed structure of GNNs, all the nodes share the same node update role, which is shown in Fig. \ref{fig:network_node}. To be specific, the update rule for the $n$-th layer at  node $v_{i,k}$ is given by   
\begin{align} \label{eq:Phiin}
\textbf{f}_{i,k,k^{'}}^{n} &= \Psi_{i}^{n}\left(\textbf{x}_{i,k^{'}}^{n-1}\right), \forall v_{i,k^{'}}\in \mathcal{N}(v_{i,k}),  \\
\textbf{m}_{i,k}^{n} &=  \Phi\left(\left\{\textbf{f}_{i,k,k^{'}}^{n} |  \forall v_{i,k^{'}}\in \mathcal{N}(v_{i,k})\right\}\right), \\ \label{eq:Omegain}
\textbf{x}_{i,k}^{n} &= \Omega_{i}^{n}\left(\textbf{m}_{i,k}^{n}, \textbf{x}_{i,k}^{n-1}\right),
\end{align}
where $\mathcal{N}(v_{i,k})$ is the set of neighbour nodes of $v_{i,k}$, $\Psi_{i}^{n}(\cdot)$ is a multi-layer perceptron (MLP) used to extract information from the neighbour node feature vectors of the last layer, $\Phi(\cdot)$ is the aggregation function that aggregates information from the neighbour nodes and  $\Omega_{i}^{n}(\cdot)$ is the combining function that combines the aggregated information $\textbf{m}_{i,k}^{n}$ with its own information $\textbf{x}_{i,k}^{n-1}$, which is parameterized by a MLP. Note that in each GNN,  $\Psi_{i}^{n}(\cdot)$ and $\Omega_{i}^{n}(\cdot)$ share the same weights for all the nodes, which enables the proposed distributed
machine learning algorithm to be generalized to an IRS-enhanced cell-free network with varying numbers of users. The aggregation function is usually chosen to be a permutation-invariant function. In this paper, we set $\Phi(\cdot)$  as an element-wise max function $F_{\text{max}}(\cdot)$ following \cite{GNN1}. After $N$ layers of graph convolutions, the node feature vectors contain the information within its $N$-hop neighborhood nodes and learn to handle inter-user interference.

\subsubsection{Output}
Each node feature vector of the $N$-th layer is followed by a fully connected layer (FCL) to output the beamforming vector or IRS reflection coefficient vector. As neural networks only support real-valued output,  for user node $v_{i,k}, \forall i\in \mathcal{I}, k\in \mathcal{K}$, a FCL with $2M$ units outputs a real-valued vector $\textbf{w}^{'}_{i,k} \in \mathbb{R}^{2M\times 1}$. Then a normalization layer (NL) is employed in each GNN. In particular, GNN $i$ converts each $\textbf{w}^{'}_{i,k}, \forall k\in \mathcal{K}$, to a complex-valued vector $\textbf{w}^{''}_{i,k}\in \mathbb{C}^{M\times 1}$ and performs the following steps:
\begin{align}
\textbf{W}_{i}^{''} &= \left[\textbf{w}^{''}_{i,1}, \textbf{w}^{''}_{i,2},\cdots,\textbf{w}^{''}_{i,K}\right] \in \mathbb{C}^{M\times K},   \\
\textbf{W}_{i} &= \sqrt{P_{\mathrm{max}}}\frac{\textbf{W}_{i}^{''}}{\|\textbf{W}_{i}^{''}\|_F^2}, \\
\textbf{w}_{i,k} &= \left[\textbf{W}_{i}\right]_{:,k}, \forall k\in \mathcal{K}.
\end{align} 

For IRS node $v_{1,0}$, a FCL with $2L$ units outputs a real-valued vector $\textbf{v}^{'} \in \mathbb{R}^{2L\times 1}$. To satisfy the unit modulus constraint, $\textbf{v}^{'}$ is followed by the following NL:
\begin{align}
\textbf{v}_{l} = \frac{\textbf{v}^{'}_{l}}{\sqrt{\left(\textbf{v}^{'}_{l}\right)^{2} + \left(\textbf{v}^{'}_{l+L}\right)^{2}}} + j \frac{\textbf{v}^{'}_{l+L}}{\sqrt{\left(\textbf{v}^{'}_{l}\right)^{2} + \left(\textbf{v}^{'}_{l+L}\right)^{2}}}, \forall l\in \mathcal{L}.
\end{align}

\subsection{Offline Training}
\label{sec:offline training}
\begin{figure}[!t]
\centerline{\includegraphics[width= 3.4in]{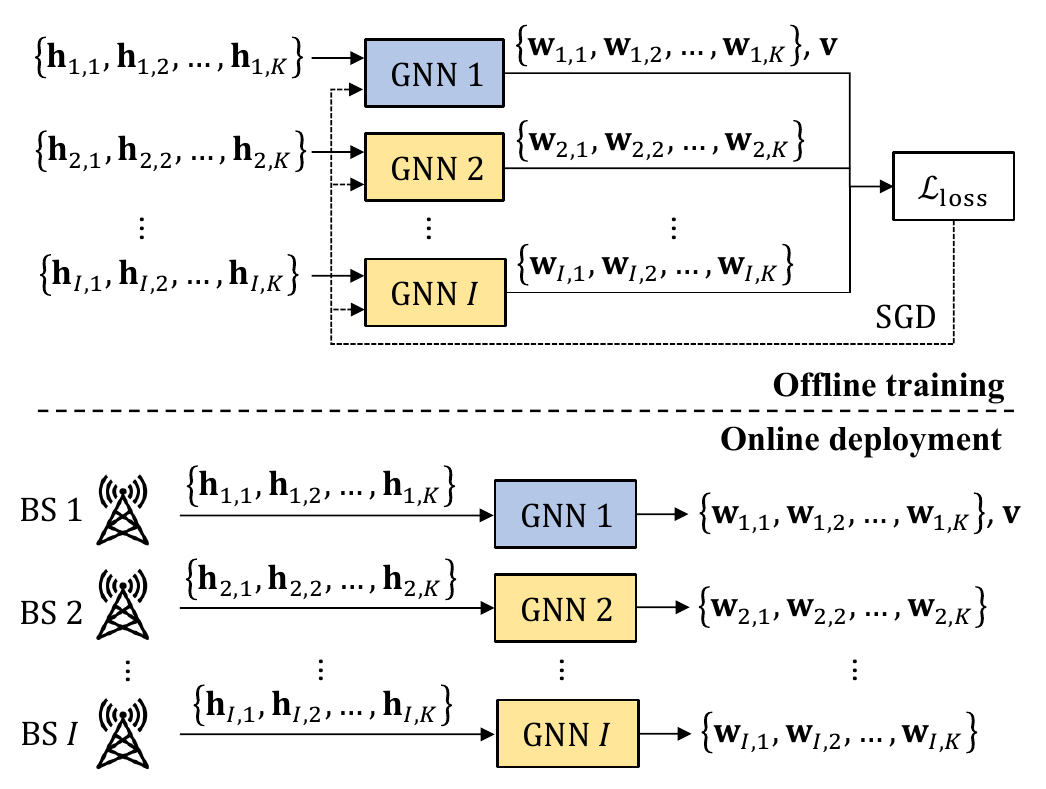}}
\caption{The proposed framework of the distributed machine learning approach.}
\label{fig:training}
\end{figure}

We adopt a centralized-training-distributed-deployment algorithm framework, as shown in Fig. \ref{fig:training}. The training phase is performed at the CPU, which has access to the global CSI information. The weight vectors of the GNNs are updated {in an unsupervised manner} according to the following loss function:
\begin{align}
\mathcal{L}_{\mathrm{loss}} = -\frac{1}{T}\sum_{t=1}^{T}\sum_{k\in\mathcal{K}}  R_{k}(\textbf{w}_k^{t}, \textbf{v}^{t}),
\end{align}
where $T$ is the number of training samples, and $\textbf{w}_k^{t}$ and $\textbf{v}^{t}$ are the beamforming vectors and IRS reflection coefficient vector determined by the GNNs for the $t$-th training sample, respectively. The GNNs learn to minimize the loss function using stochastic gradient descent (SGD). Since the loss function is the opposite value of the optimization objective function, the sum user rate increases as the loss function decreases. The calculation of the loss function requires the global CSI information, which enables the GNNs {to} learn to deal with interference from other BSs. {Noting that the training phase is completed offline, the computational complexity is less a concern.} 

\subsection{Online Deployment}
\label{sec:online deployment}
Different from the centralized training, the online deployment is performed in a  fully distributed manner. To do so, each BS updates its local GNN by downloading the well trained weight vectors from the CPU. After obtaining the local CSI information, each BS can determine its local beamforming vectors directly using its local GNN.  The IRS reflection coefficient vector is determined by the GNN of BS $1$. In contrast to the conventional optimization algorithms that require the global CSI information, the well trained distributed machine learning algorithm requires no CSI exchange between the BSs and the CPU.

\subsection{Complexity Analysis}
As the training phase is performed offline, we focus on the computational complexity of the online inference phase.  For the input layer, the computational complexity of the average operation at the IRS node over $K$  user input features is $\mathcal{O}(2KM(L+1))$. We denote the dimensions of $\Psi_{i}^{n}$ and $\Omega_{i}^{n}$ by $\mathcal{H}_{1,i}^{n}=\{h_{q_{1,i}^{n}}|q_{1,i}^{n}=1,2,\cdots,Q_{1,i}^{n}\}$ and $\mathcal{H}_{2,i}^{n}=\{h_{2,i}^{n}|q_{2,i}^{n}=1,2,\cdots,Q_{2,i}^{n}\}$, respectively, where $Q_{1,i}^{n}=|\mathcal{H}_{1,i}^{n}|$ and $Q_{2,i}^{n}=|\mathcal{H}_{2,i}^{n}|$. Each neuron of the MLPs uses an activation function. Accordingly, the computational complexity of the graph convolutional  layers is 
$\mathcal{O}\left(\sum_{i=1}^{I}K\left(C_{i} + \sum_{n=1}^{N} h_{Q_{1,i}^{n}}\right)+C_{1}-\sum_{i=2}^{I}\sum_{n=1}^{N}h_{Q_{1,i}^{n}}\right)$,
where $C_{i}=\sum_{n=1}^{N}\sum_{j\in\{1,2\}}\sum_{q_{j,i}^{n}=1}^{Q_{j,i}^{n}}\!\left(h_{q_{j,i}^{n}\!-\!1} \!+\! 1\right)h_{q_{j,i}^{n}}$.  For the output layer, the computational complexity is $\mathcal{O}\left(2MK\left(\sum_{i=1}^{I}h_{2,i}^{N} + I\right) + 2L\left(h_{2,1}^{N} + 1\right)\right)$. The total computational complexity is the sum of the computational complexity of the input layer, graph convolutional layers and output layer.

\section{Simulation Results}
\label{sec:simulation}
In this section, we provide performance evaluation for the proposed distributed machine learning algorithm in comparison with the benchmark methods. 

\subsection{Simulation Setup}
\begin{figure}[!t]
\centerline{\includegraphics[width= 3.2in]{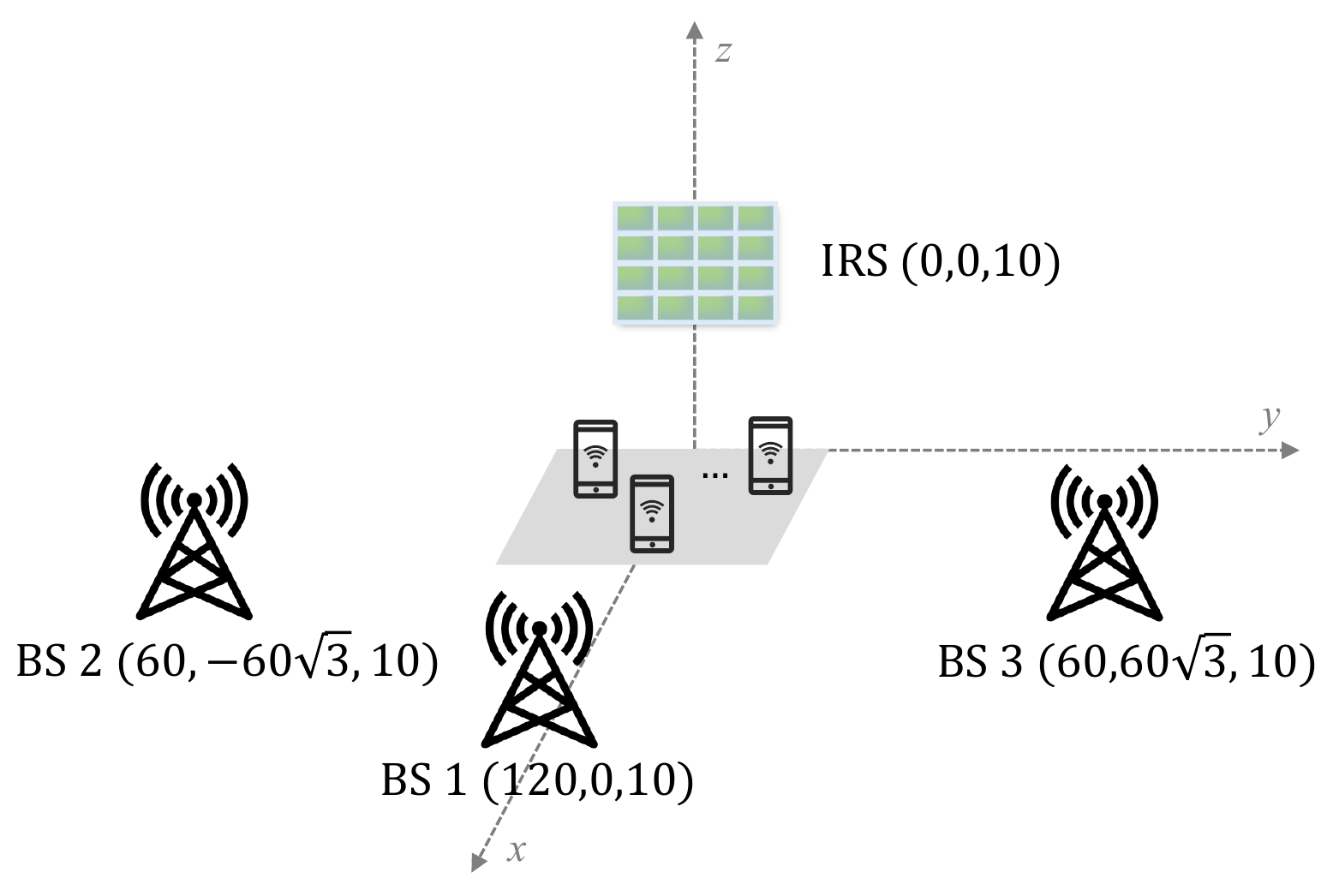}}
\caption{Simulation setup for the IRS-enhanced cell-free MIMO network.}
\label{fig:simulation}
\end{figure}

In the simulations, we consider an IRS-enhanced cell-free MIMO network as illustrated in Fig. \ref{fig:simulation}, where there are $I=3$ BSs cooperatively  serving the users. The coordinates of the BSs are ($120, 0, 10$) m, ($60, -60\sqrt{3}, 10$) m and ($60, 60\sqrt{3}, 10$) m, respectively.  The IRS is located at ($0, 0, 10$) m. The users are randomly distributed in the $xy$-plane with $x\in [0, 20]$ m and  $y\in [-20, 20]$ m to achieve generalization to user locations. The BSs are assumed be uniform linear arrays parallel to the $x$-axis. The IRS is assumed to be a uniform square array parallel to the $yz$-plane with $\sqrt{L}$ elements per column and $\sqrt{L}$ elements per row. The noise power is set {to} $\delta^2 = -90$ dBm.

The large-scale path loss in dB is computed by 
\begin{align}
	\beta = \beta_{0} - 10\alpha\text{log}_{10}\left(\frac{d}{d_0}\right),
\end{align}
where $\beta_{0}=-30$ dB is the path loss at the reference distance, $\alpha$ is the path-loss exponent, $d$ is the length of the transmission link and $d_{0}=1$~m is the reference distance. Due to the effect of blockages, the path-loss exponent of BS-user channels is set {to} $\alpha_{\text{BU}}=3.75$. IRSs are usually deployed at appropriate locations to provide line-of-sight (LOS) links, and thus the path-loss exponents of BS-IRS channels and IRS-user channels are set {to} $\alpha_{\text{BI}}=\alpha_{\text{IU}}=2.2$. The small-scale fading of BS-user channels is modelled as Rayleigh fading, which is composed of uncorrelated $\mathcal{CN}(0,1)$ elements. For IRS-assisted links, the small-scale fading is assumed to be Rician fading. To be specific, the small-scale fading of the channel from BS $i$ to the IRS is given by
\begin{align} 
\widetilde{\textbf{G}}_{i} = \sqrt{\frac{\kappa}{1+\kappa}}\widetilde{\textbf{G}}_{i}^{\text{L}} + \sqrt{\frac{1}{1+\kappa}}\widetilde{\textbf{G}}_{i}^{\text{NL}},
\end{align}
where $\kappa$ is the Rician factor, which is set {to} 10, $\widetilde{\textbf{G}}_{i}^{\text{L}}$ and $\widetilde{\textbf{G}}_{i}^{\text{NL}}$ are the LOS component and non-line-of-sight (NLOS) component of $\widetilde{\textbf{G}}_{i}$, respectively. $\widetilde{\textbf{G}}_{i}^{\text{NL}}$ is Rayleigh fading, which consists of uncorrelated $\mathcal{CN}(0,1)$ elements. When a plane wave is impinging on the RIS from the azimuth angle $\varphi\in [-\pi, \pi]$ and elevation angle $\theta\in [-\frac{\pi}{2}, \frac{\pi}{2}]$, the array response vector {is given} by
\begin{align} \label{eq:arva}
\textbf{a}(\varphi, \theta) = \textbf{a}_{z}(\theta)\otimes \textbf{a}_{x}(\varphi, \theta),
\end{align}
where 
\begin{align} 
\textbf{a}_{z}(\theta)&=\left[1, e^{\frac{j2\pi d_{\text{I}}}{\lambda}\mathrm{sin}(\theta)}, \cdots,e^{\frac{j2\pi(\sqrt{L}-1)d_{\text{I}}}{{\lambda}}\mathrm{sin}(\theta)} \right]^{T}, 
\end{align}
and
\begin{align} 
\textbf{a}_{x}(\varphi, \theta)&=\left[1, e^{\frac{j2\pi d_{\text{I}}}{\lambda}\mathrm{sin}(\varphi)\mathrm{cos}(\theta)}, \cdots,e^{\frac{j2\pi(\sqrt{L}-1)d_{\text{I}}}{{\lambda}}\mathrm{sin}(\varphi)\mathrm{cos}(\theta)} \right]^{T}
\end{align}
 are the array response vectors with respect to the vertical direction and   the horizontal direction, respectively, in which $\lambda$ is the wavelength and $d_{\text{I}}$ is the IRS element spacing. In the simulations, we consider the half-wavelength element spacing, i.e., $d_{\text{I}}=\frac{\lambda}{2}$. Similarly, when a plane wave is impinging on the BS from
the azimuth angle $\psi\in [-\pi, \pi]$ and elevation angle $\phi\in [-\frac{\pi}{2}, \frac{\pi}{2}]$, the array response
vector is given by 
\begin{align} \label{eq:arvb}
    \textbf{b}(\psi, \phi)=\left[1, e^{\frac{j2\pi d_{\text{B}}}{\lambda}\mathrm{cos}(\psi)\mathrm{cos}(\phi)}, \cdots,e^{\frac{j2\pi(M-1)d_{\text{B}}}{{\lambda}}\mathrm{cos}(\psi)\mathrm{cos}(\phi)} \right]^{T},
\end{align}
where $d_{\text{B}}$ is the BS antenna spacing {and set to} half-wavelength, i.e., $d_{\text{B}}=\frac{\lambda}{2}$. Applying (\ref{eq:arva}) and (\ref{eq:arvb}), $\widetilde{\textbf{G}}_{i}^{\text{L}}$ can be expressed as
\begin{align} 
\widetilde{\textbf{G}}_{i}^{\text{L}}=
\textbf{a}(\varphi_{\text{BI},i}, \theta_{\text{BI},i})\textbf{b}^{H}(\psi_{\text{BI},i}, \phi_{\text{BI},i}),
\end{align}
where $\varphi_{\text{BI},i}$ and $\theta_{\text{BI},i}$ denote the azimuth and
elevation angles of arrival (AoA) from BS $i$ to the IRS, respectively, and $\psi_{\text{BI},i}$ and $\phi_{\text{BI},i}$ denote the azimuth and
elevation angles of departure (AoD) from BS $i$ to the IRS, respectively.

The small-scale fading of the channel from user $k$ to the IRS is given by
\begin{align}
\widetilde{\textbf{f}}_{k} = \sqrt{\frac{\kappa}{1+\kappa}}\widetilde{\textbf{f}}_{k}^{\text{L}} + \sqrt{\frac{1}{1+\kappa}}\widetilde{\textbf{f}}_{k}^{\text{NL}},
\end{align}
where $\widetilde{\textbf{f}}_{k}^{\text{L}}$ and $\widetilde{\textbf{f}}_{k}^{\text{NL}}$ are the LOS component and NLOS component of $\widetilde{\textbf{f}}_{k}$, respectively. $\widetilde{\textbf{f}}_{k}^{\text{NL}}$ is Rayleigh fading consisting of uncorrelated $\mathcal{CN}(0,1)$ elements and $\widetilde{\textbf{f}}_{k}^{\text{L}}$
is expressed as
\begin{align}
\widetilde{\textbf{f}}_{k}^{\text{L}} = \textbf{a}(\varphi_{\text{UI},k}, \theta_{\text{UI},k}),
\end{align}
where $\varphi_{\text{UI},k}$ and $\theta_{\text{UI},k}$ are the azimuth and
elevation AoA from user $k$ to the IRS. 

\subsection{GNN Setup}
As illustrated in Section \ref{sec:structure of gNNs}, each BS is equipped with a GNN.  We set $N=2$, i.e., each GNN has 2 layers of graph convolutions. In our simulations, all the MLPs share the same network structure. In particular, the information extraction function $\Psi_{i}^{n}$ in (\ref{eq:Phiin}) and the combining function $\Omega_{i}^{n}$ in (\ref{eq:Omegain}) are respectively parameterized by a $1600\times 800$ MLP.  We adopt the leaky rectified linear unit (LReLU) as the activation function \cite{Lu_2020}. For input $X$ and output $Y$, the LReLU function is given by 
\begin{align}
 Y  = 
 \left\{\!
\begin{array}{ll}
X, & {X\ge 0},
\\
0.1X, &  {X<0}.
\end{array} \right.
\end{align}
The SGD is implemented by Adam optimizer with an initial learning rate of 0.01. We adopt the learning rate decay strategy. {For every} 100 training steps, the learning rate is reduced by a factor of 0.995. 
In each training epoch,
60000 groups of random user locations are used as training samples with
a batch size of 600. That is, each training epoch has 100 training steps. We terminate the training when the number of training epochs reaches 2000 or the loss function on the verification data set does not {decrease} by more than 10 training epochs.

\subsection{Numerical Results}
\label{sec:numerical_results}
We present the simulation results of the proposed distributed machine learning algorithm in comparison with the following benchmark methods:
\begin{itemize}
\item Global ZF + PA: Represent the baseline solution developed in Section \ref{sec:baseline}.

\item Global ZF: The ZF beamformer is computed centrally at the CPU using the global CSI with equal power allocation. The  reflection coefficients of the IRS are configured randomly.

\item Local ZF: Each BS computes its ZF beamformer according to its own local CSI with equal power allocation. The  reflection coefficients of the IRS are configured randomly.

\item Maximum ratio transmission (MRT) \cite{7827017}: MRT precoding matrix at each BS is equal to its channel matrix, and thus no CSI exchange is required. Equal power allocation is adopted. The  reflection coefficients of the IRS are configured randomly.
\end{itemize}
All presented simulation results are averaged results over 500 independent channel realizations.

\begin{figure}[t]
\centerline{\includegraphics[width=3.2in]{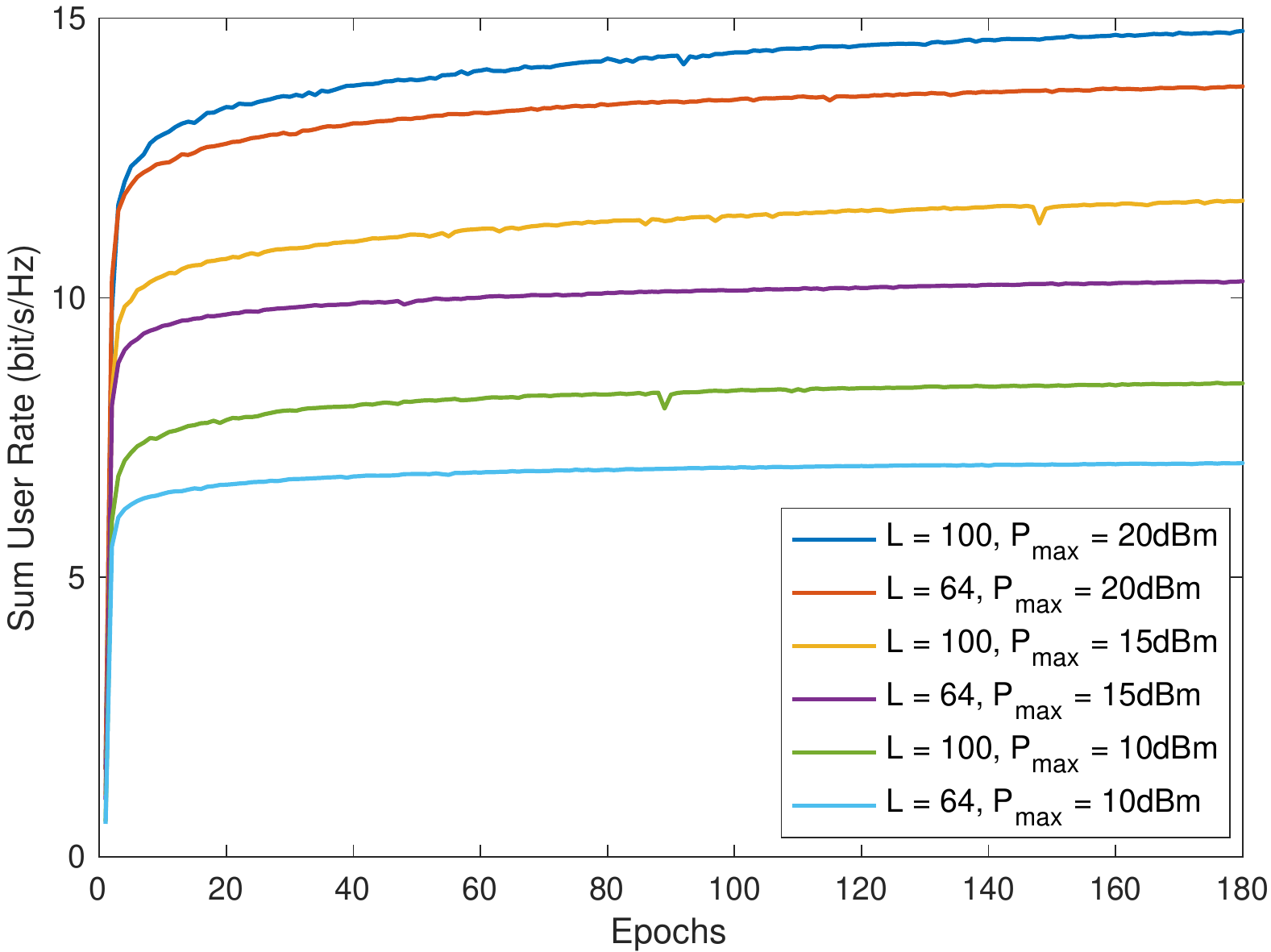}}
\caption{Sum user rate versus training epochs with $M=4$ and $K=3$.}
\label{fig:training_result}
\end{figure}

Fig. \ref{fig:training_result} shows the convergence of the proposed distributed machine learning algorithm. It is clear that the proposed algorithm can effectively learn how to improve the sum user rate. The transmit power $P_{\text{max}}$ and the number of IRS elements $L$ have effects on the convergence rate. Generally, larger $P_{\text{max}}$ and $L$ lead to slower convergence. Moreover, it is observed that for all the parameter settings, only 20 training epochs can achieve a satisfactory sum user rate.

 \begin{figure}[t]
\centerline{\includegraphics[width=3.2in]{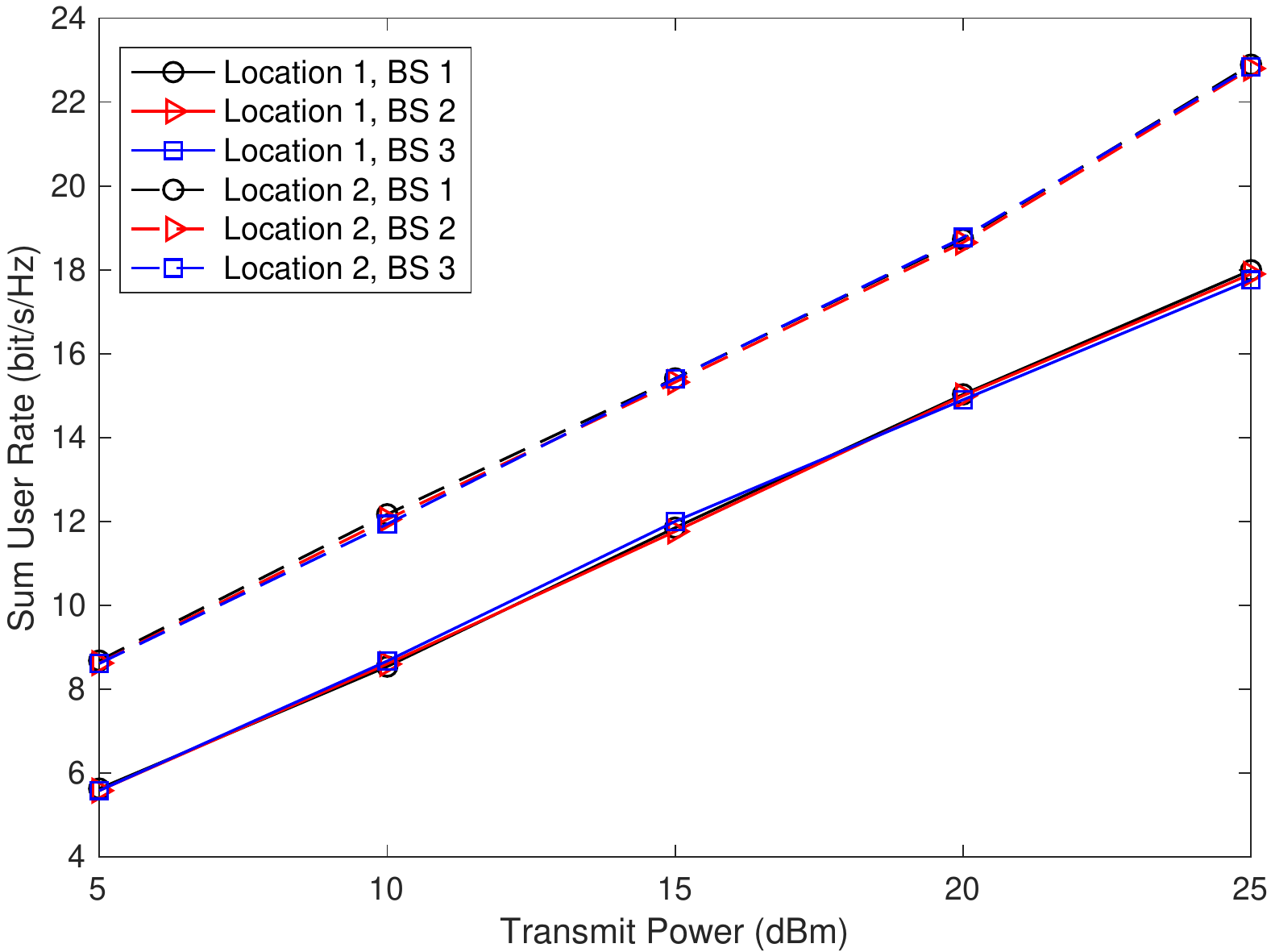}}
\caption{The effect of different BSs determing the IRS reflection coefficients  with $M=4$, $K=3$ and $L=100$.}
\label{fig:location}
\end{figure}

Fig. \ref{fig:location} presents the impact  of different BSs determining the IRS reflection coefficients on the sum user rate. ``Location 1'' denotes the default coordinates of the BSs. For ``Location 2'', we set the coordinates of BS 1-3 to ($30, -30\sqrt{3}, 10$) m, ($90, 0\sqrt{3}, 10$) m and ($60, 60\sqrt{3}, 10$) m, respectively. We can see that determining the IRS reflection coefficients by an arbitrary BS can achieve the same sum user rate, regardless of the BS coordinates.

\begin{figure}[t]
\centerline{\includegraphics[width=3.2in]{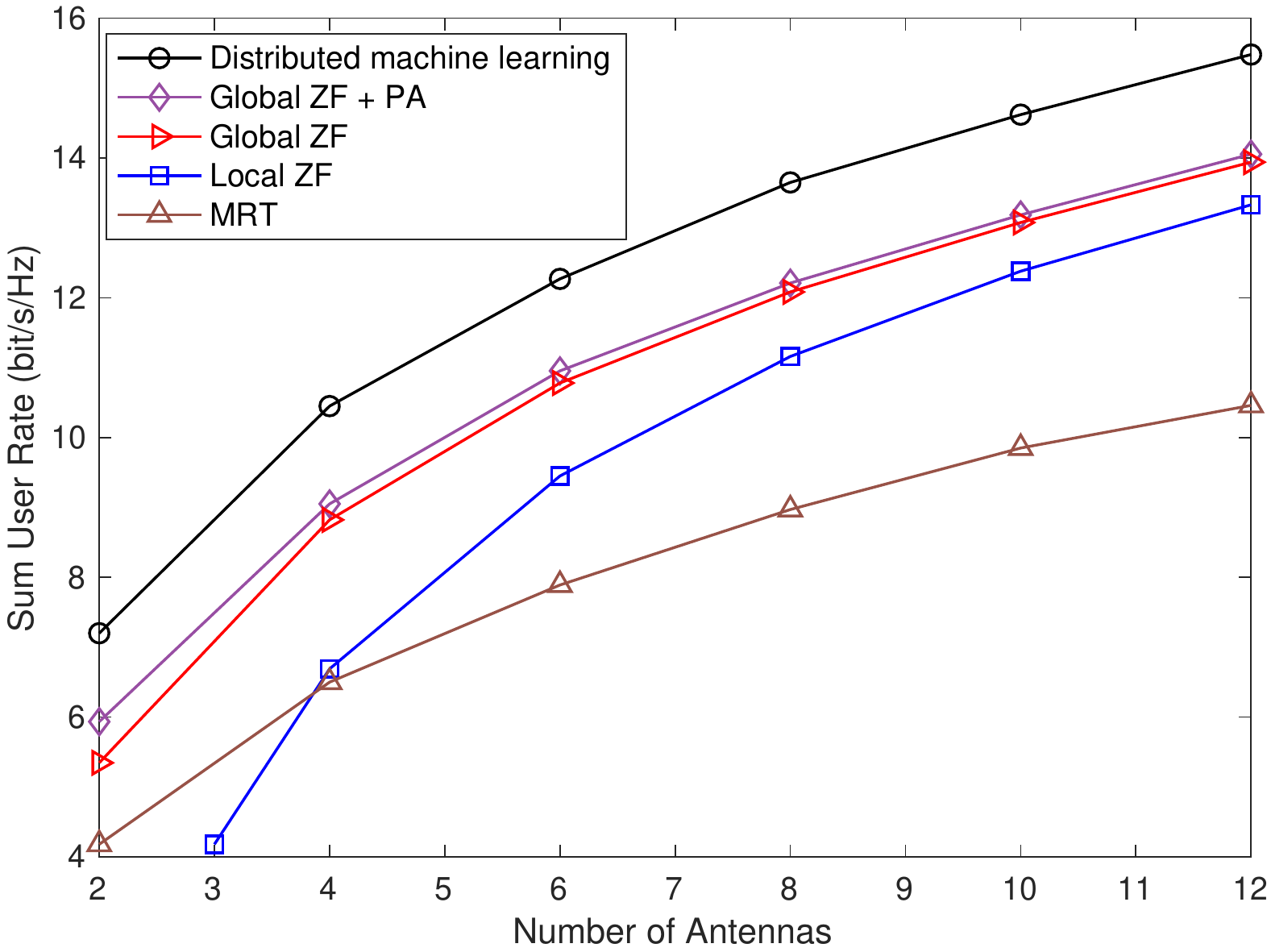}}
\caption{Sum user rate versus the number of antennas $M$ with $K=3$, $L=64$ and $P_{\mathrm{max}}=15$ dBm.}
\label{fig:antenna}
\end{figure}

In Fig. \ref{fig:antenna}, we show the effect of the number of antennas $M$ on
the sum user rate. First of all, we can see that the proposed distributed machine learning algorithm outperforms all the benchmark methods, which verifies its effectiveness. To be specific, without any CSI exchange, the proposed algorithm achieves a performance gain of 21.3\% and 10.2\% compared to the global ZF + PA method when $M=2$ and $M=12$, respectively. This demonstrates that the optimization of IRS  reflection coefficients can significantly improve the sum user rate. The curve of local ZF starts from $M=3$ due to the constraint $M\ge K$. Although the beamforming vectors can be computed locally, the local ZF method suffers from a low sum user rate when $M$ is small. Despite simple and distributed implementation, the MRT method cannot achieve a satisfactory sum user rate. 
  
\begin{figure}[t]
\centerline{\includegraphics[width=3.2in]{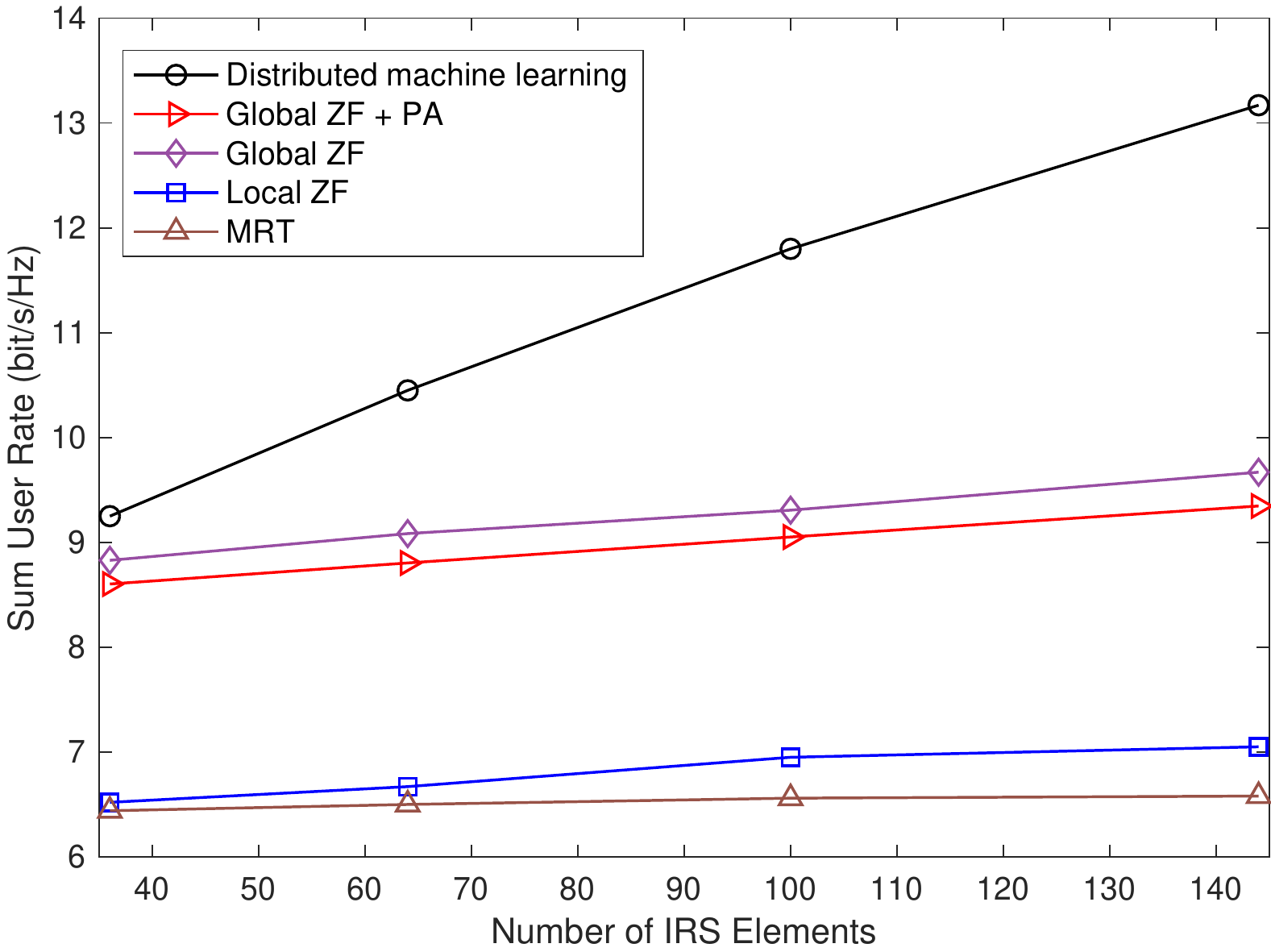}}
\caption{Sum user rate versus the number of IRS elements $L$ with $M=4$, $K=3$ and $P_{\mathrm{max}}=15$ dBm.}
\label{fig:element}
\end{figure}

In Fig. \ref{fig:element}, the sum user rate versus the number of IRS elements $L$ is displayed. For all the methods considered, the sum user rate monotonically increases with $L$, indicating that the deployment of a larger IRS can improve the sum user rate. Specifically, the proposed distributed machine learning algorithm achieves 42.38\% sum user rate gain when $L$ increases from $36$ to $144$. It is worth noting that the performance gap between the proposed distributed machine learning algorithm and the benchmark methods significantly increases  with $L$, which implies the importance of optimizing  reflection coefficients at the IRS. 

\begin{figure}[t]
\centerline{\includegraphics[width=3.2in]{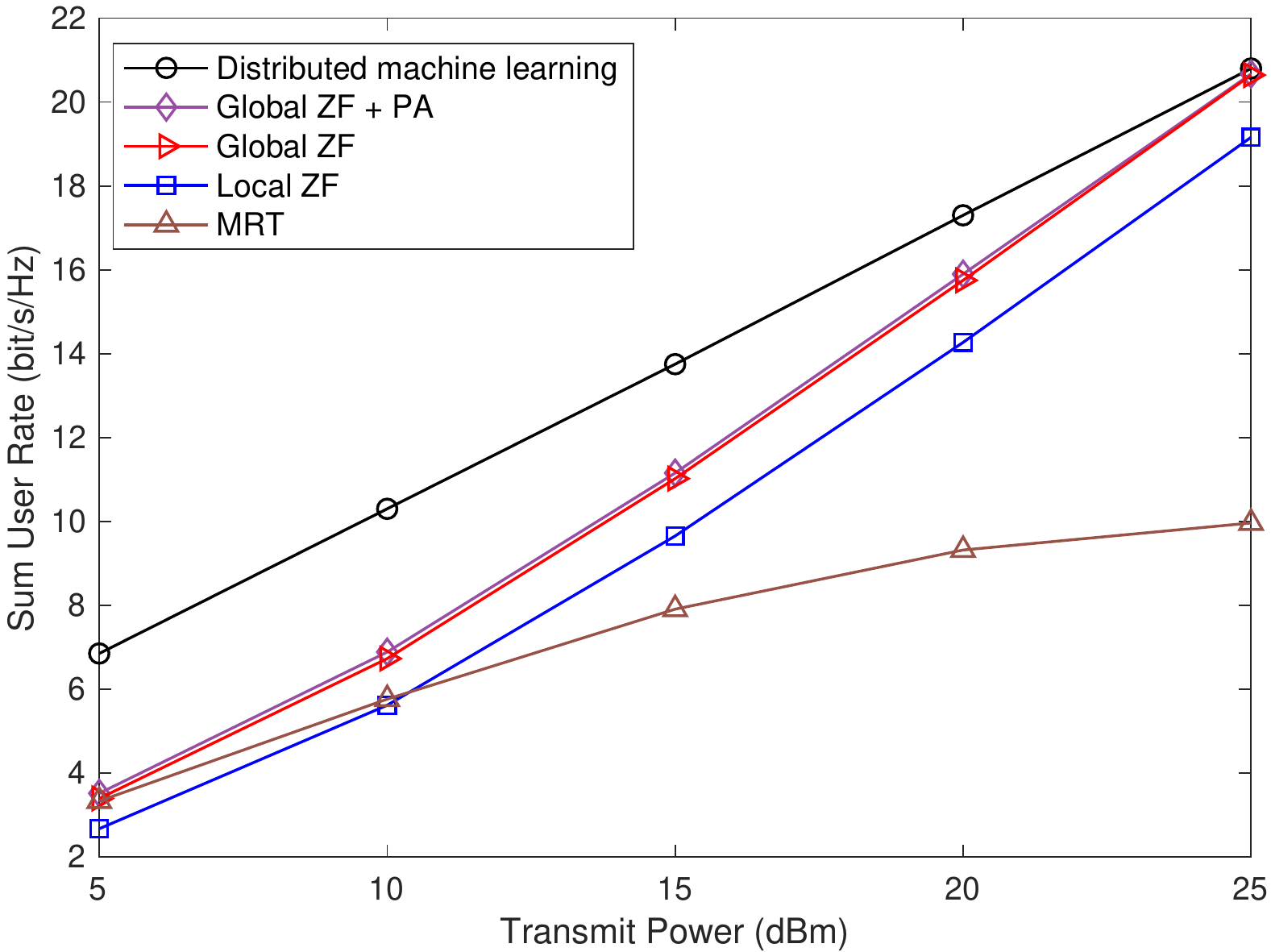}}
\caption{Sum user rate versus the maximum transmit power $P_{\text{max}}$ with  $M=6$, $K=3$ and $L=100$.}
\label{fig:power}
\end{figure}

In Fig. \ref{fig:power}, we present the effect of transmit power $P_{\text{max}}$ on the sum user rate. Notably, the proposed distributed machine learning algorithm can achieve a higher sum user rate than the benchmark methods for all the values of $P_{\text{max}}$.  Additionally, we observe that the global ZF + PA, global ZF and local ZF methods benefit more from from the increase of $P_{\text{max}}$ than other methods. When $P_{\text{max}}=25$ dBm, they can achieve a similar sum user rate as the  the proposed distributed machine learning algorithm. On the other hand, they suffer from a poor performance for a low SNR ($P_{\text{max}}$). 

 \begin{figure}[t]
\centerline{\includegraphics[width=3.2in]{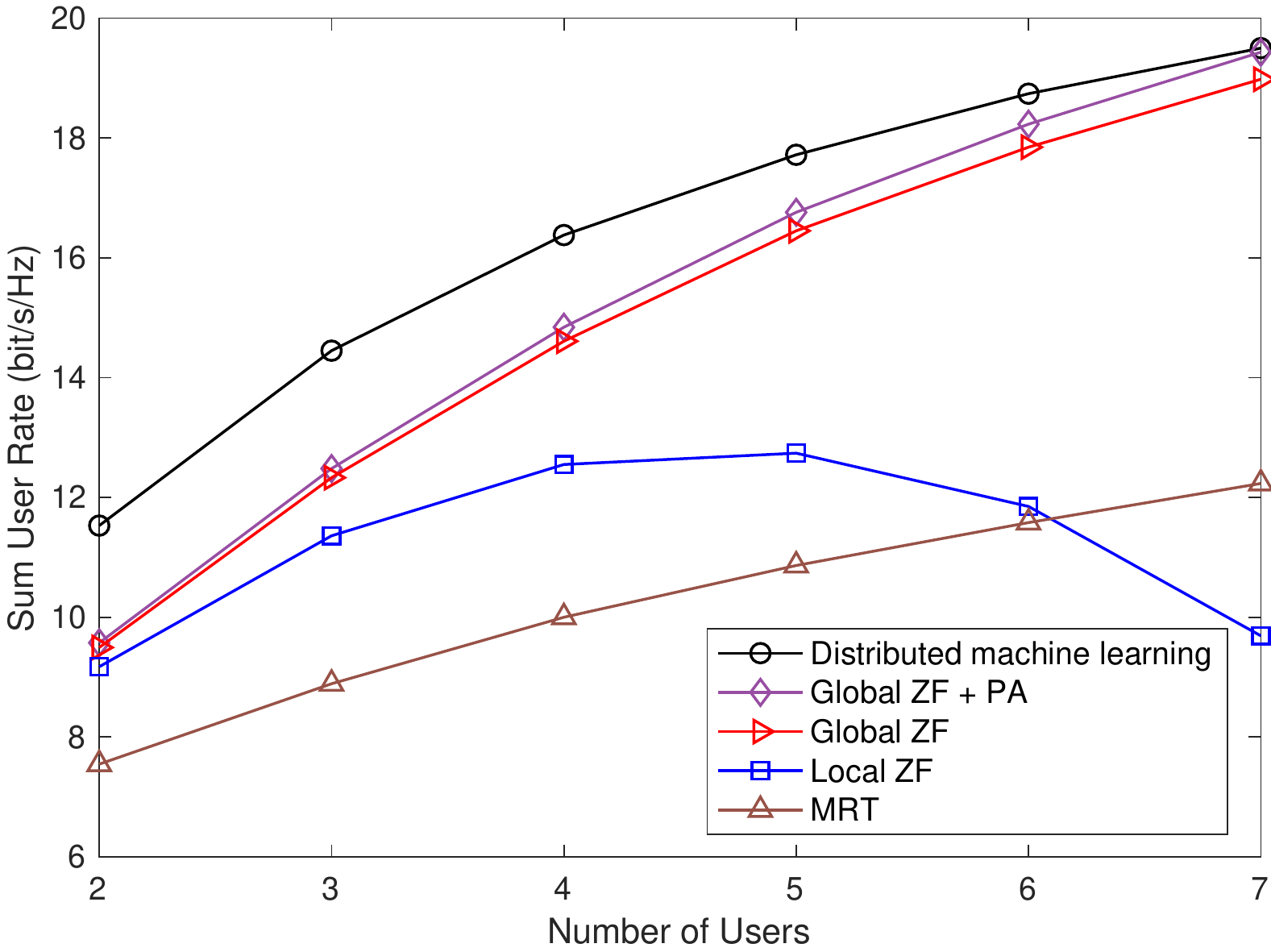}}
\caption{Sum user rate versus the number of users $K$ with $M=8$, $L=100$ and $P_{\mathrm{max}}=15$ dBm.}
\label{fig:user}
\end{figure}

In Fig. \ref{fig:user}, we test the generalization capability of the proposed distributed machine learning algorithm
to different numbers of users. We train the GNNs with $K=5$ and apply the trained GNNs to different numbers of users during the test phase. From Fig. \ref{fig:user}, it can be seen that the proposed  algorithm has a good generalization capability to different numbers of users. The proposed algorithm outperforms the benchmark methods in terms of sum user rate, while the gap between the proposed algorithm and the benchmark methods decreases with the increasing  number of users.
This is because for a large number of users, the global ZF method can better eliminate inter-user interference with the global CSI. In contrast, for a large number of users, the distributed method cannot achieve the optimal performance due to the limited interference cancellation capability of a single BS.

Table \ref{tab:run_time} compares the sum user rate, computational time and signaling exchange with the benchmark methods, where the proposed distributed machine learning algorithm is referred to as ``DML''. We run the algorithms for 1000 times on the same platform, a 4 core Intel Core i7 CPU with 2.2 GHz base frequency and 16GB memory, and record the average sum user rate in bit/s/Hz and the average computational time in milliseconds.  Since the  global ZF + PA method requires CVX programming, we implement it using MatLab. All other methods are written in Python 3.8. 
We can see that the computational time of the global ZF + PA method is significantly longer than the other methods. Note that the computational time recorded for the proposed DML algorithm is the total time of all the GNNs. In practical, the GNNs are implemented in parallel on different BSs, and the computational time per GNN is about 0.63 ms and 1.28 ms for $K=3$ and  $K=6$, respectively. With comparable computational time, the proposed DML algorithm achieves a significantly higher sum user rate than the global ZF and local ZF methods. Moreover, the DML algorithm requires no CSI and signaling exchange for determining the transmit and reflect  beamforming.
The only signaling exchange is the reflection coefficients determined by the GNN that need to be transferred to the IRS.

\begin{table*}[t]
  \centering
  \caption{Comparison of Sum User Rate, Computational Time, and Signaling Exchange ($M=8$, $L=100$, $P_{\mathrm{max}}=15 \ \mathrm{dBm}$)}
    \begin{tabular}{c|c|c|c|c|c|c}
    \toprule
    \multirow{2}[4]{*}{Algorithm} & \multicolumn{2}{c|}{K=3} & \multicolumn{2}{c|}{K=6} & \multicolumn{1}{c|}{\multirow{2}[4]{*}{CSI Exchange}} & \multicolumn{1}{c}{\multirow{2}[4]{*}{\tabincell{c}{Signaling Exchange \\ (beamforming)}}} \\
\cmidrule{2-5}          & \multicolumn{1}{p{7.335em}|}{Sum User Rate} & Computational Time & \multicolumn{1}{p{7.335em}|}{Sum User Rate} & Computational Time &       &  \\
    \midrule
    DML   & 14.45 & 1.89  & 18.74 & 3.85  & 0     & 2L \\
    \midrule
    Global ZF + PA & 12.5  & 811   & 18.25 & 1314  & 2IMK  & 2IMK \\
    \midrule
    Global ZF  & 12.31 & 0.28  & 17.7  & 0.46 & 2IMK  & 2IMK \\
    \midrule
    Local ZF & 11.36 & 0.29  & 11.84 & 0.48  & 0     & 0 \\
    \bottomrule
    \end{tabular}%
  \label{tab:run_time}%
\end{table*}%

\section{Conclusions}
\label{sec:conclusions}
In this paper, we have investigated an IRS-enhanced downlink cell-free MIMO network in harsh propagation environments. Aiming to maximize the sum user rate, we have formulated an optimization problem to jointly design the transmit beamforming vectors at the BSs and the reflect beamforming vector at the IRS.
We have exploited the recent advances of machine learning and proposed a novel fully distributed machine learning algorithm to solve the optimization problem. The proposed algorithm deploys a GNN at each BS and requires no CSI exchange between the BSs and the CPU. Simulation results reveal that the proposed distributed machine learning algorithm outperforms the benchmark methods in terms of sum user rate and has a low computational complexity. Compared with the conventional machine learning structures, the proposed algorithm has a better generalization capability to different numbers of users. In our future work, we will address the problem of imperfect CSI and explore the deployment of multiple IRSs.



\end{document}